\documentstyle[pre,aps,epsfig,multicol]{revtex}

\begin{document}

\title{ Collective motion and solid-liquid-type transitions in 
vibrated granular layers }
\author{\small{ \it{
Nicol\'as Mujica and Francisco Melo} \\ 
Departamento de F\'\i sica de la Universidad de Santiago de
Chile, \\ 
Av. Ecuador 3493, Casilla 307 Correo 2 Santiago-Chile}\\}

\vskip 1 cm

\maketitle

\begin{abstract}

From pressure and surface dilation measurements, we show that a
solid-liquid-type transition occurs at low excitation frequencies in
vertically vibrated granular layers.  This transition
precedes subharmonic bifurcations from flat surface to standing 
wave
patterns, indicating that these 
waves are in fact associated with the fluid like behavior of the
layer. In the limit of  high excitation frequencies, we show that a new
kind of subharmonic waves can be distinguished.  These waves do 
not 
involve any lateral transfer of grains within the layer and 
correspond to excitations for which the layer slightly bends 
alternately in time and space.  These bending waves have very
low amplitude and we observe them in a vibrated two-dimensional 
layer of photoelastic particles.

\author{\parbox{430pt}{\vglue 0.3 cm \small PACS numbers: 
46.10.+z, 83.10.Ji, 83.70.Fn}}
\maketitle

\end{abstract}

\vskip 1.5 cm

\begin{multicols}{2}

\section{ Introduction}
\begin{figure}[t!]
\centering
\leavevmode
\epsfxsize=14cm 
\epsfysize=3.5cm 
\epsfbox{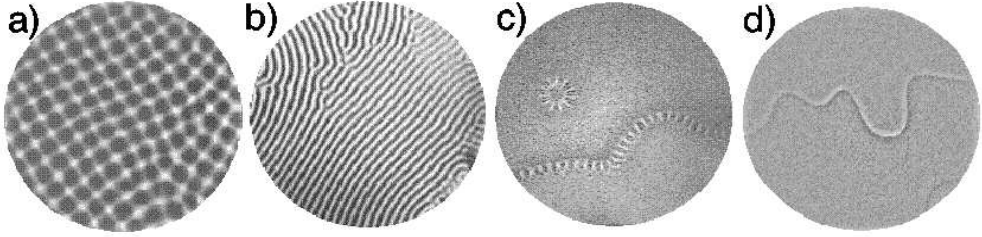}
\vspace{0.3cm}
\caption{\protect\small Snapshots of the layer state at differents 
conditions 
obtained in a large cell.  a) Squares, $f \sim {20}$ Hz, $\Gamma= 
3.0$.  b) Stripes, $f \sim {60}$ Hz,
$\Gamma= 3.0$.  c) Flat with kinks at $f \sim {25} $ Hz, $\Gamma= 
4.7$. d) Flat with kinks at $f \sim {60}$ Hz,
$\Gamma= 4.7$. } 
\label{fig1}
\end{figure}

In a 
driven granular system, such as a vibrated layer composed of 
macroscopic particles, energy is 
dissipated by inelastic  grains
collisions and by friction.  Also, for realistic situations, the 
typical energy of a grain is many 
orders of magnitude greater 
than $k_BT$, so the temperature does not play an important 
role in the dynamics of these materials.   However, it is commonly 
observed that granular matter behaves like  
solids,  liquids or even gases depending on the energy injection 
and 
energy dissipation rates \cite{jnb96}.  More precisely,
the competition between these quantities determines the 
residual velocity fluctuations which  in turn play
the role of thermal fluctuations in these materials. 
From this view point, surface waves excited by vertical vibrations 
provide
one of the most striking examples in which granular materials 
act like a fluid.  For a granular layer, it was well established 
that 
parametric waves can be observed whenever the dimensionless 
acceleration $\Gamma=A(2\pi f)^{2}/g$ exceeds a critical value 
\cite{mus94} (here $g$ is the acceleration of gravity, $A$ is the 
amplitude of the vibrating surface, and $f$ is the frequency of 
the 
driving force).  The primary instability resulted from a flat 
layer to a pattern of squares or 
stripes, depending on both $f$ and the particle diameter, 
$d$.  As an illustration we present two typical snapshots of these 
kind of waves in Fig.~\ref{fig1} a and b.  It was found that the 
crossover
frequency $f_{d}$ at which the square to stripe transition occurs 
is 
proportional to $d^{-1/2}$.  In addition, this scaling was 
qualitatively 
understood in terms of the ratio of kinetic energy injected into 
the 
layer to the potential energy required to raise a particle a 
fraction 
of its diameter.  At low 
$f$, the horizontal mobility should be high since the layer 
dilation is large \cite{haff,ums96}, whereas at large $f$, 
the mobility should be low since layer dilation is small 
\cite{mm98}.  We notice, however, that these waves are the 
corresponding 
hydrodynamic surface modes of the layer since they require a 
transfer of mass to be sustained \cite{mus94,ums96,mm98}.  

In this paper, we report on measurements of the pressure due to 
the 
layer-container collision and the surface and bulk dilation of the 
layer.  We 
show that there exists a critical $\Gamma$ for which the flat 
layer 
undergoes a phase transition.  This critical $\Gamma$ is smaller 
than the 
critical one at which waves appear.  At low $f$ this transition 
is a 
solid-liquid type, at intermediate $f$ only a heating up of the 
surface 
layer is observed, whereas at high $f$ a compaction transition is 
detected.  Thus,
our results show that  granular surface waves are naturally linked 
to 
the fluid like behavior since they are 
observed only when the energy input per 
particle is enough to induce a minimal dilation of the flat layer.  
At high $f$ and for the same critical value of 
$\Gamma$ at which hydrodynamic waves appear, bending waves 
are detected instead.  In 
this regime, our measurements show that the mobility of grains is 
almost completely suppressed 
in both 
the bulk and the free surface of the layer.  In this case, a set 
of
experiments conducted with large photoelastic particles allow us 
to observe 
these waves as an alternation of bright and dark zones that 
oscillate in time at half of the forcing frequency.  In 
addition, we show for the low frequency regime that when $\Gamma$ 
is
increased to a value close to $4.6$, an inverse 
transition from a fluid to a compact layer is 
observed. This transition is responsible for the flat with kinks 
state reported in previous works \cite{mus94} (See 
Fig.~\ref{fig1} c and d).  A brief summary of our results has  
been given in ref. \cite{mm98}.

This article is organized as follows:  section II is devoted 
to 
the description of our experimental setup. 
In section III, which is the main part of this article,  we 
present 
our pressure measurements and we establish the basis to obtain 
density profiles and the bulk dilation from the experimental data.  
In section IV we present surface dilation measurements and we 
discuss how they link to the bulk dilation measurements.  The 
energy dissipation rate in the vibrated layer is 
also estimated as a function of the excitation frequency.  Section 
V 
is concerned with the study of the very low amplitude regime of 
surface waves and its connection with the compact state of the 
layer.  Finally a brief discussion is given.

\end{multicols}

\begin{multicols}{2}

\section{Experimental Setup}

In this article we present two sets of experiments. In the first 
set, we measure simultaneously the pressure resulting from the 
collisions of the layer with the vibrating plate and the surface
particle dilation. From the pressure measurements, we obtain 
information
on the bulk dilation and density (see section III). The surface dilation is determined through the
normalized reflected intensity of the surface layer (See 
discussion below). 

\begin{figure}[t!]
\centering
\leavevmode
\epsfxsize=9cm 
\epsfysize=6.3cm 
\epsfbox{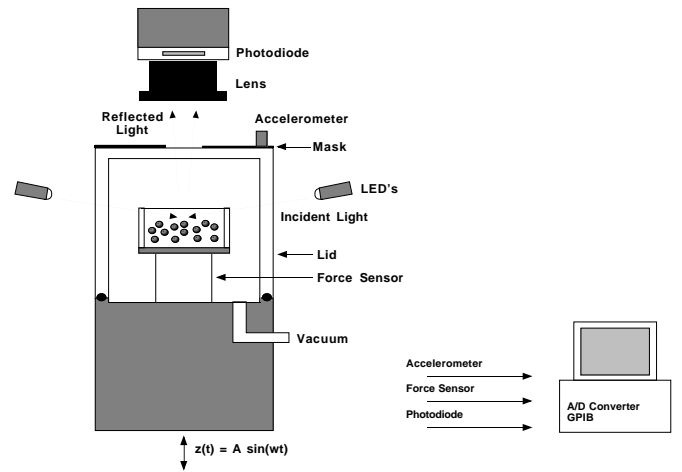}
\vspace{0.3cm}
\caption{\protect\small A schematic drawing of the apparatus showing
the cylindrical cell, the location of the pressure sensor, and the
setup for reflectivity measurements.} 
\label{fig2}
\end{figure}

In fig.~\ref{fig2} we present a 
schematic drawing of the experimental 
setup. In this experiment, a thin layer of $0.106-0.125$ mm 
diameter bronze particles, 15 particles deep, is placed at the 
bottom of a $40$ mm diameter and $25$ mm height cylindrical 
container.  The container's wall is Lucite while the base is 
aluminum to reduce electrostatic effects.  The container is 
mounted on a high frequency response pressure sensor (PCB. 
Model 208A11) 
which is driven by an electromechanical vibration exciter. The 
resulting acceleration is measured to a resolution of $0.01g$.  A 
second Lucite cylinder is used as a lid for the whole system 
allowing evacuation of the container to less than $0.1$ Torr; at 
this 
value volumetric effects of the gas are negligible \cite{pdb95}. 
The 
surface of the layer is illuminated at low angle ($20^{0}$ respect 
to 
the horizontal) by an 
array of  $18$ LEDs organized in a $10$ cm diameter ring.  The 
reflected light from the surface layer is focused by a lens of 
$28$ mm focal length on a flat photodiode of $25$ mm$^{2}$ area.  
The whole system is automatically run by a Power PC computer 
equipped with A/D and GPIB boards.  

With this set-up, the measured light is proportional to both the 
incident light and the reflectivity coefficient of the bronze 
particles 
$R$ ($R\sim 
{0.6}$).  We notice that only a small fraction $s$, which is about  
$5$ 
percent of the surface of a single particle, reflects light in the 
direction of the 
solid angle of the camera (in the experiment, lens aperture angle 
is 
about $12^{0}$).  The intensity $I$, measured by the photodiode, 
can be then taken  
proportional to the surface density or more precisely to the 
number 
of particles within the first layer.  Furthermore, $I$ relates to
surface dilation $\delta_{s}$  as $I/I_{0} \approx {d^{2}/(d + 
\delta_{s})^{2}}$, where $I_{0}$ is a reference intensity for 
which we 
take $\delta_{s} = 0$ \cite{init_cond}. We neglect multiple 
reflections,  since their dominant contribution is proportional to 
both $R^{2}$ and $s^{2}$, where $s^{2}$  represents the 
probability 
of having 
a secondary reflection within the solid angle of the camera. Thus, 
by taking the incident light in a small enough angle, we insure 
that, to a first approximation, only the first layer of particles 
contribute to $I$.

The second set of experiments consists of vibrating a two 
dimensional layer of photoelastic particles and taking images of 
the layer motion with a high speed CCD camera. The 
goal is to observe the very low amplitude waves found in the 
high frequency regime. A cell made of two glass plates $400$ mm 
wide by $100$ mm high was mounted on the moving platform of 
our vibrator system.  The gap  between the plates is controlled 
by spacers of 
varying thickness to a resolution of $0.05$ mm. Images with 
resolution of $256 \times 256$ pixels are captured at 
rates of $1200$ frames per second by a Hisis $2002$ CCD 
camera.
Images are obtained by transmission using parallel light with
incidence perpendicular to the cell.  Transmitted light is filtered through a  
polarizer whose axis is perpendicular to direction of polarization 
of the incident light. More details are given in section V.

\section{Pressure measurements and density profiles}

\subsection{Experimental procedures and collision model}

\begin{figure}[th]
\centering
\leavevmode
\epsfxsize=8.5cm 
\epsfysize=4.26cm 
\epsfbox{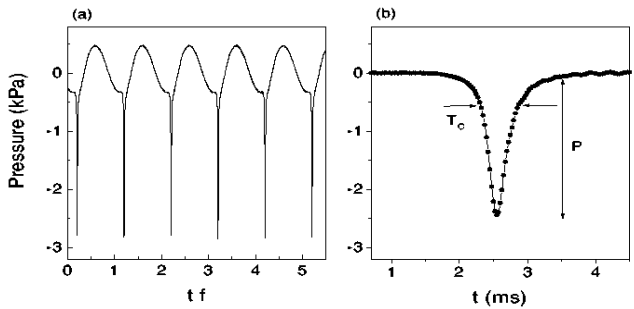}
\vspace{0.3cm}
\caption{\protect\small (a) Time evolution of the pressure signal 
$P(t)$ for $f=40$ Hz and $\Gamma=2.2$. In (b) we present the 
averaged collision peak obtained from $P(t)$, and we show the 
definition of the maximum pressure $P$. The collision time 
$T_c$ is the width of the peak at certain height.}
\label{fig3}
\end{figure}

In Fig. \ref{fig3}a, we present a typical pressure signal 
$P(t)$ as a function of time for $f=40$ Hz and $\Gamma = 2.2$. 
This signal is composed of a sinusoidal component, corresponding 
to the force required to accelerate the cell, and of a peak 
sequence, due to the layer-plate collisions. We are interested in 
the shape of the peak during the collision so it is necessary to 
subtract the sinusoidal component. By this procedure we obtain 
a number of pressure peaks, typically about 15. As an illustration, we 
show in Fig. \ref{fig3}b the average peak obtained from 
the peaks presented in Fig. \ref{fig3}a.
To characterize the peaks of a sequence like the one presented we 
take the maximum value  $P$ and the width $T_c$, which is a 
measure of the collision time. 
More precisely these quantities are obtained as the averages of 
the sequence $\{ P^{(k)},T_c^{(k)} \}$, where $k$ is an integer 
that 
indicates the collision number, i.e.  $P=\left < P^{(k)} \right >$ 
and 
$T_c=\left < T_c^{(k)} \right >$ where $\left < \, \right >$ 
denotes 
the sequence average. For the data in the Fig. \ref{fig3}, $P 
= 
2.45 \pm 0.03$ kPa and $T_c = 0.52 \pm 0.01$ ms (In this case 
$T_c$ is taken as the width at a pressure equal to a quarter of 
$P$). 

Now, due to the impulsive nature of the layer plate collision, we 
write the following scaling relation
\begin{equation}
P \sim \frac{M}{A_p} \frac{V_c}{T_c} 
\label{p_vc_tc}
\end{equation}
where $M$ is the total mass of the layer, $A_p$ is the active 
surface of the plate and $V_c$ is the layer-plate relative 
velocity  
at the collision. If the collision takes place at time $t=\bar{t}$ 
and 
is completely inelastic then $V_c = V_p(\bar{t}) - V_b(\bar{t})$, 
where $V_p$  and $V_b$ are the plate and layer velocity 
respectively. Thus, in practice, we approximate the collision 
velocity by the one predicted by the completely inelastic ball 
model. We note that this model assumes that the layer losses 
immediately all its energy and takes the velocity of the moving 
plate. This approximation is valid since previous experiments have 
shown, by measuring the flight time of the layer, that the motion 
of the center of mass of the layer follows the motion of an 
inelastic 
ball \cite{mus94,ums97}.  

We remark that in our experiments, even close to 
$\Gamma\sim{1}$, the vertical average bulk dilation of 
the layer,  $\delta_b$, is never strictly zero.  This is concluded because the 
collision time is not determined by the Hertz theory. Indeed, 
experiments 
performed by E. Falcon {\it et al} \cite{falcon} have shown that, 
in 
the case of a column of $N$ particles in contact ($\delta_b =0$) 
colliding vertically with a fixed plate, the collision time is 
$T_c = 
(N-1) T_q + \tau_1$, where $N$ is the number of particles, $T_q$ 
is the time duration of the momentum transfer from one particle 
to another, and $\tau_1$ is the collision time between two 
particles predicted by Hertz theory \cite{landau_elasticity}.
Now, for the particles used in our experiments the orders of 
magnitude of these times are $T_q \approx \tau_1 \approx 10^{-
6}$ s \cite{tiempos}. Then, the predicted collision time for our 
three dimensional layer should be of order $T_c \approx 10^{-5}$ 
s. This order of magnitude is never observed in our experiments, 
where for $\Gamma$ 
close to 1 and for a wide range of $f$ the minimum collision time 
is of order 
$T_c \approx {5 \times 10^{-4}}$ s.  
The difference between what is expected from the Hertz 
prediction for $\delta_b =0$ and the 
experimental values for $T_c$ can be explained if the layer is 
slightly dilated at the collision; a local dilation of 1 percent  
($\delta_b/d\sim{0.01}$) drastically changes the collision regime 
from Hertz contact to ballistic collisions.  Although this effect 
has 
interesting consequences on sound propagation, it does not affect 
the center of mass motion of the layer.  Thus, for $\Gamma > 1$ 
we write the total dilation of the layer as $\Delta = V_c T_c$. If 
in 
addition we assume that dilation is homogeneous, i.e. $\Delta = N 
\delta_b$, where $N$ is the number of layers, we can write 
equation (\ref{p_vc_tc}) as
\begin{equation}
P \sim \frac{M}{A_p} \frac{V_c^2}{N \delta_b} 
\label{p_vc2_delta}
\end{equation}
Thus, our procedure to measure the local bulk dilation $\delta_b$ 
is: we fit $P$ versus $V_c/T_c$ to obtain the numerical factor of 
equation (\ref{p_vc_tc}) and we then use equation  
(\ref{p_vc2_delta}) to obtain the value of $\delta_b$.

\begin{figure}[t]
\centering
\leavevmode
\epsfxsize=6.7cm 
\epsfysize=10cm 
\epsfbox{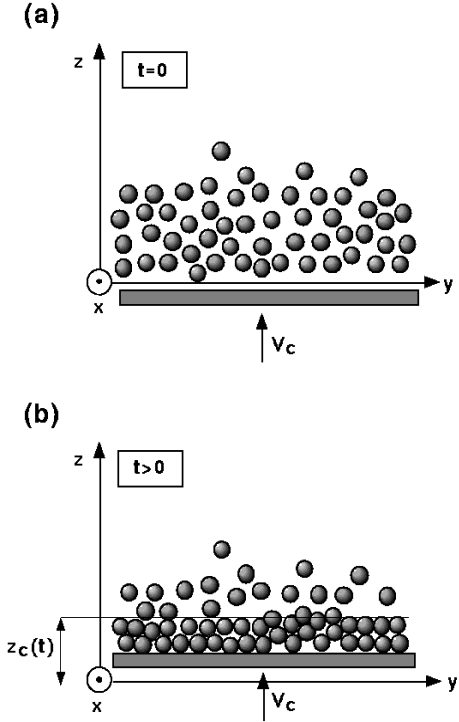}
\vspace{0.3cm}
\caption{\protect\small Schematic of the collision. Fig. (a) shows 
the instant $t=0$ at which the plate colides with the layer. The 
velocity of the collision is $V_c$ and the density of the layer 
before 
the collision is $\rho(z)$; Fig. (b) shows how the layer 
accumulates over the moving plate at an instant $t>0$. We also 
present the definition of  $z_c(t)$ as the position of the 
compresion front relative to the origin of the fixed frame 
$(x,y,z)$. 
In all the calculations the effect of gravity is neglected, since 
experimentally $g T_c \ll V_c$.}
\label{fig4}
\end{figure}

In the following, we will  relate the layer density to the pressure signal 
$P(t)$.  Fig. \ref{fig4} presents a schematic view of a 
collision. We consider that the plate collides with the granular layer 
of 
density  $\rho(z)$ at a velocity $V_c$ at time $t=0$. Then, the 
layer is initially fixed in space and the plane $(x,y,z=0)$ of 
the 
reference frame $(x,y,z)$ 
coincides with the inferior layer just before the collision (See 
Fig. 
\ref{fig4}a). We also notice that the mass of the plate 
$M_p$ is much larger than the mass $M$ of the layer ($M/M_p \sim 
0.01$) so we 
neglect any velocity change of the plate due to momentum 
transfer.  We also neglect the variation of the relative velocity 
due 
to the action of gravity since experimentally we observe that $g 
T_c <<  V_c $. Thus, the force exerted on the plate due to de 
momentum transfer is

\begin{equation}
F(t) \approx \frac{dm(t)}{dt} V_c
\label{ec_fuerza}
\end{equation}
Here $m(t)$ is the mass that has collided with the plate at time 
$t$.  
In the next we need to link $dm(t)/dt$ to  $\rho(z)$,  
measured in the fixed frame $(x,y,z)$. To do this we define the 
compression front $z_c(t)$ as the height (relative to the fixed 
frame) of the mass that has been deposited on the plate at time $t 
$ (see Fig. \ref{fig4}b). So, the mass $m(t)$ accumulated 
on the plate becomes
\begin{equation}
m(t) = A_p \int \limits_0^{z_c(t)} {\rho( z)dz}
\end{equation}
whose derivative with respect to time is
\begin{equation}
\frac{dm(t)}{dt} = A_p \rho(z_c(t)) \frac{dz_c(t)}{dt}
\label{ec_dmdt}
\end{equation}
Now, at time $t$ the plate has moved a distance $V_c t$ from its 
initial position, and the height of the layer with respect to the 
plate can be written
\begin{equation}
h(t) \equiv \frac{m(t)}{A_p\rho_o} = \frac{1}{\rho_o} \int 
\limits_0^{z_c(t)} {\rho(z)dz}
\end{equation}
where $\rho_o$ is the density of the layer in the compact state.  
We then obtain
\begin{equation}
z_c(t) = V_c t + \frac{1}{\rho_o} \int \limits_0^{z_c(t)} 
{\rho(z)dz}
\label{ec_xc_implicita}
\end{equation}
Differentiating this equation respect to time, we obtain the density 
evaluated at the location of the compression front as a function 
of its velocity $\dot{z}_c(t)$
\begin{equation}
\rho(z_c(t)) = \rho_o \left (  1 - \frac{V_c}{\dot{z}_c(t)} \right 
)
\label{rho_t}
\end{equation}
Using this result in equation (\ref{ec_dmdt}), the relation
(\ref{ec_fuerza}) for the force becomes
\begin{equation}
\frac{F(t)}{A_p \rho_o V_c} = \dot{z}_c(t) - V_c
\label{ec_F2}
\end{equation}
Integrating this equation between $0$ and $t$ gives us 
an expression for the compression front as a function of time
\begin{equation}
z_c(t) = V_c t + \frac{1}{A_p \rho_o V_c} \int \limits_0^t 
{F(\bar{t}) 
d\bar{t}}
\label{ec_xc_P}
\end{equation}
In summary, from the pressure signal $P(t)=F(t)/A_p$ we deduce the 
time 
evolution of the compression front  $z_c(t)$. Also, we use 
equation 
(\ref{rho_t}), which tells us about the time evolution of density at 
the compression front.  Since the time $t$ enters as a simple 
parameter we can  obtain the density as a function of $z_c$, right 
before the collision.

Finally, we notice that both equations (\ref{rho_t}) and
(\ref{ec_F2}) are laws for the conservation of mass and momentum 
through
the compression front. It is possible to deduce from them the
velocity of the compression front and the pressure in the 
compressed
part of the layer as implicit functions of $\rho(z_c(t))$, 
\begin{eqnarray*}
\dot{z}_c(t) &=& \frac{\rho_o V_c}{\rho_o - \rho(z_c(t))} \\
\frac{F(t)}{A_b} &=& \frac{\rho_o \rho(z_c(t)) V_c^2}{\rho_o - 
\rho(z_c(t))}
\end{eqnarray*}
These expressions are  generalizations of those obtained by 
Goldshtein {\it
et al}  \cite{gsg96} for the propagation
of a shock wave through an homogeneous layer of inelastic 
particles.

Before presenting our experimental results we will show that 
with equation (\ref{ec_xc_P}) we can check the momentum 
conservation of the collision.  Evaluating it at time $t=T_c$ 
(here 
$T_c$ is the total collision time) we obtain 
\begin{equation}
z_c(T_c) = V_c T_c + \frac{1}{A_p \rho_o V_c} \int \limits_0^{T_c} 
{F(\bar{t}) d\bar{t}}
\end{equation}
Since by definition $z_c(T_c) = H + V_c T_c$, where $H$ is the 
layer
thickness in the compact state and $V_c T_c$ is the total 
displacement of the 
plate during the collision, and $\rho_o = M/H A_p$, we find
\begin{equation}
\int \limits_0^{T_c} {F(\bar{t}) d\bar{t}} = M V_c 
\label{cons_momentum}
\end{equation}

\subsection{Experimental results}

We begin this section by presenting results concerning the 
momentum conservation
relation (\ref{cons_momentum}). Experimentally we have checked it 
for a
wide range of parameters, namely $1 < \Gamma < \Gamma_w \approx 
2.8$ and $35$ Hz $ < f < 350 $ Hz.  The main point is that, in our 
experiment, the velocity
$V_c$ calculated from the completely inelastic ball model is a very 
good
approximation.  As the internal degrees of freedom of the
layer are
excited one expects that this approximation becomes less 
accurate. However, this only occurs in the wave regime
where the take off velocity is reduced and then 
both the collision velocity and the flight time are smaller than
predicted \cite{mus94}. 

There are other experimental effects that should be considered.  For 
instance,
friction on the cell walls can tranfer momentum to the layer. Another
posibility,  
much less probable, is the transfer of momentum to the walls by the
formation of dynamical arcs.
Therefore, we expect that the correct form of the momentum
conservation, taking into account all possible sources of errors, to be
\begin{equation}
\int \limits_0^{T_c} {F(\bar{t}) d\bar{t}} = M_{eff} V_c 
\label{cons_momentum2}
\end{equation}
where $M_{eff}$ is an effective mass, and $V_c$ is the
velocity of collision calculated from the completely inelastic 
ball model. From this model, we know that
$V_c = g F(\Gamma)/f$, were $F(\Gamma)$ is a nonanalytic function 
of
$\Gamma$, which is found numerically \cite{ums97}. From our data
we find then that the scaling $\int \limits_0^{T_c} {F(\bar{t})
d\bar{t}} \sim V_c$ is very well verified as a function of both 
$\Gamma$ and
$f$.  As expected, we find that $M_{eff}$ is independent of both $\Gamma$
and $f$ and is slightly lower than $M$.

In order to complete our description, it is necessary to estimate 
the
granular density in the compact state 
$\rho_o$. This quantity is in principle a dynamical variable, in 
the sense
that it depends on how we compact our layer.  However, we
notice that the important parameter in (\ref{ec_xc_P}) is $A_p 
\rho_o
= M/H$.  Taking in to account the previous discussion about the
effective mass of the layer and that $H \approx 1.7$ mm, we obtain 
$A_p \rho_o \approx 4.2$ kg/m.

\begin{figure}[p]
\centering
\leavevmode
\epsfxsize=12cm
\epsfysize=9.2cm
\epsfbox{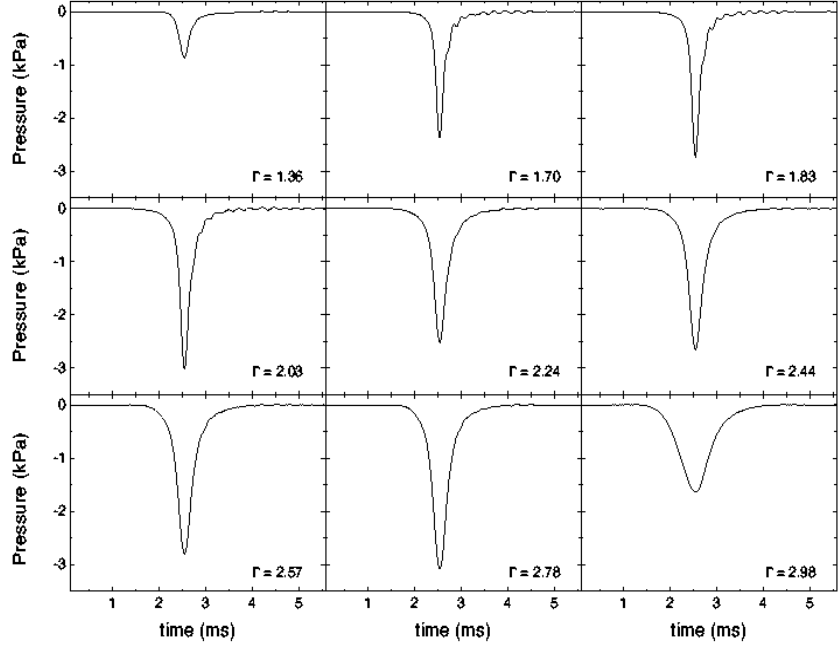}
\vspace{0.3cm}
\caption{\protect\small Averaged pressure peaks as functions of 
time for various values of  $\Gamma$ and $f=40$ Hz. For 
$\Gamma=2.98$ the layer is already in the wave regime.}
\label{fig5}
\end{figure}

Fig. \ref{fig5}  presents averaged pressure peaks for 
several values of $\Gamma$ at $f=$ 40 Hz.  Each curve is obtained 
by
averaging 15 collisions, in the same way as that for the one 
presented in Fig. \ref{fig3}b.  We notice that our analysis 
is valid for cases for which $\Gamma < \Gamma_w$, where 
$\Gamma_w \approx 2.8$ is the onset for parametric waves. Thus, 
the last pressure curve presented for $\Gamma = 2.98$ is shown 
to display the 
difference of the parametric wave state; at this value of $\Gamma$ 
the collision is quite spread out in time and the maximum 
pressure achieved is also much smaller. We will discuss 
more this point in the next section. For the other pressure curves 
presented, the intensity of the 
collision seems to be an increasing function of $\Gamma$, i.e. the 
maximum pressure $P$ increases. We understand this as the 
simple fact that the relative velocity of the collision $V_c$ is 
increasing with $\Gamma$, as it does in this region of $\Gamma$ 
in the completely inelastic ball model. What is not possible to 
explain with this model is the observation that the collision 
time $T_c$ also seems to be an increasing function of $\Gamma$. 
As discussed before, this is due to the excitation of the internal 
degrees of freedom of the granular layer, i.e,  the layer 
is  dilated.

\begin{figure}[p]
\centering
\leavevmode
\epsfxsize=12cm 
\epsfysize=9.2cm 
\epsfbox{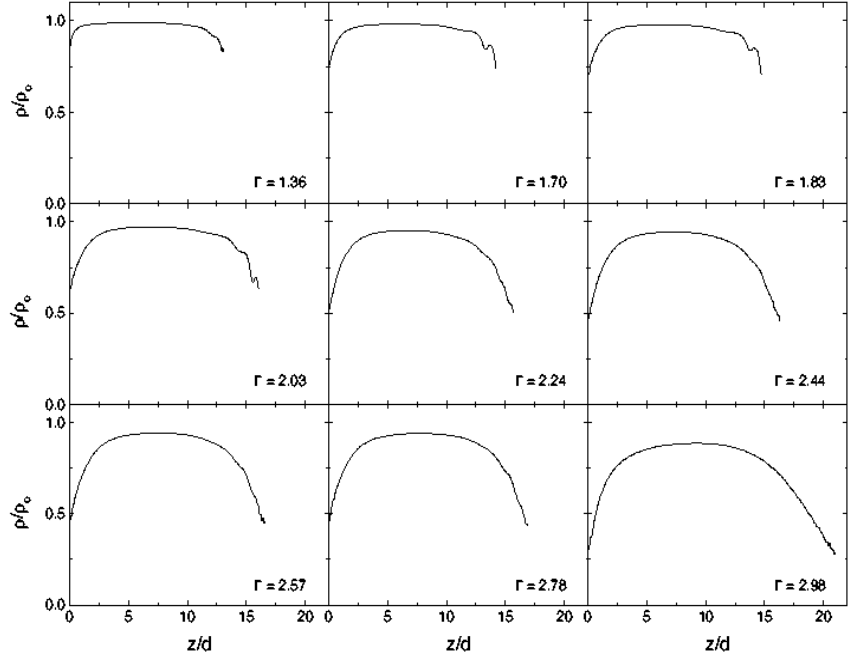}
\vspace{0.3cm}
\caption{\protect\small Normalized density, $\rho(z) / \rho_o$, as 
function of height, $z$, for various values of $\Gamma$ and $f=40$ 
Hz.}
\label{fig6}
\end{figure}

In Fig.  \ref{fig6} we present the layer density as a 
function of height for each of the curves introduced in Fig. 
\ref{fig5}.  In general, the density $\rho(z)$ is approximately 
constant in a center region and decreases toward both ends of the 
layer.  A penetration length for the dilation of the layer can be 
identified.  This length increases with $\Gamma$;  for instance, 
it is of the order of $d$ and $3d$ for $\Gamma \approx 1.4$ 
and $\Gamma \approx 2.4$ respectively.  Another fact that traces 
back the dilatance of the layer is that its thickness increases 
with  
$\Gamma$; it is $14d$ for $\Gamma \approx 1.4$ and $17d$ for 
$\Gamma \approx 2.8$.  In the limit of $\Gamma \sim 1$ we 
obtain that the density in the central part of the layer is close 
to 
$\rho_o$.

\begin{figure}[p]
\centering
\leavevmode
\epsfxsize=12cm 
\epsfysize=9.2cm 
\epsfbox{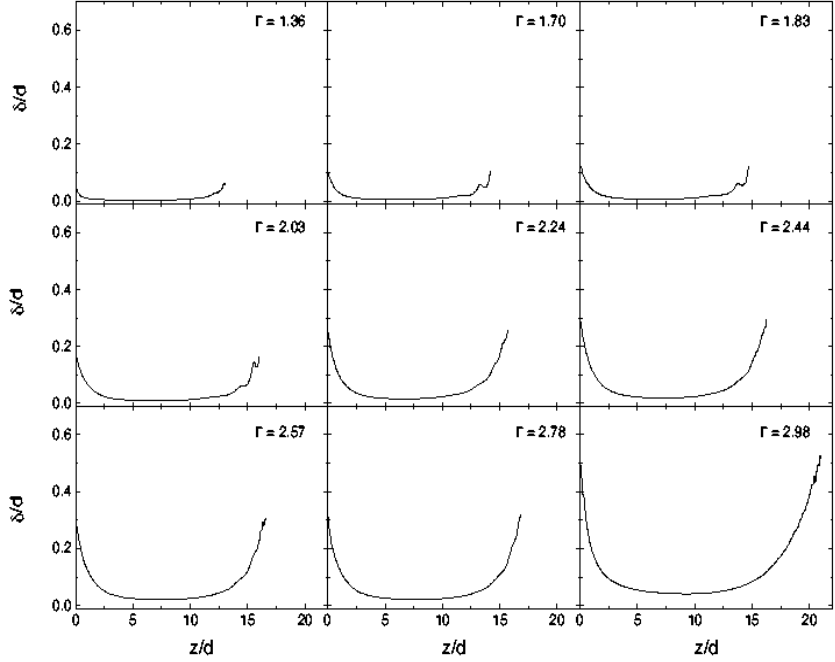}
\vspace{0.3cm}
\caption{\protect\small Normalized dilation, $\delta (z) / d$,  as 
function of height, $z$, for various values of $\Gamma$ and $f=40$ 
Hz.}
\label{fig7}
\end{figure}

Another interesting piece of information is the bulk dilation within the 
layer, $\delta$, as a function of $z$.  This quantity is linked directly 
to the density as $\rho(z) = \bar{\nu} m_o / (d+\delta)^3$, where 
$\bar{\nu}$ is an average coordination number and $m_o$ the 
particle mass \cite{haff}. Defining $\delta = 0$ for $\rho(z) = 
\rho_o$ we obtain
\begin{equation}
\frac{\delta(z)}{d} = \left ( \frac{\rho(z)}{\rho_o} \right )^{-
\frac{1}{3}} - 1 
\end{equation}
Then , it is possible to find $\delta(z)$ for all the data 
presented 
previously, which is what is shown in Fig. \ref{fig7}.  From this figure, 
we can qualitatively examine the dependence of  $\delta(z)$ on 
$\Gamma$, as we did with $\rho(z)$.  For instance, as we increase 
$\Gamma$, the top and the bottom of the layer continuously dilate 
and the extension of the dilated part increases.  The central part 
also dilates as $\Gamma$ increases; for $\Gamma \approx 1.4 $ and
$\Gamma \approx 2.4$ it is of the order of $0.004d$ and $0.02d$
respectively.  We also notice that for $\Gamma 
\approx 2.4$, the value $\delta \approx 0.1d$ is reached for $z 
\approx 2d$ and $z \approx 15d$, while the total height of the 
layer is approximately $17d$.

\end{multicols}

\begin{multicols}{2}

At this point, it seems appropriate to define a quantity as the 
vertical average of $\delta(z)$; we do this because it is 
necessary to 
link it in some way to $\delta_b$.  Therefore, we define this 
average dilation as
\begin{equation}
\left < \delta \right > = \frac{1}{H'} \int \limits_0^{H'} 
\delta(z) dz
\end{equation}
where $H'$ is the total height of the layer. Notice that $H'$ 
depends
on the excitation intensity (i.e, it depends on $\Gamma$ and $f$).

\begin{figure}[p]
\centering
\leavevmode
\epsfxsize=8.5cm 
\epsfysize=6.1cm 
\epsfbox{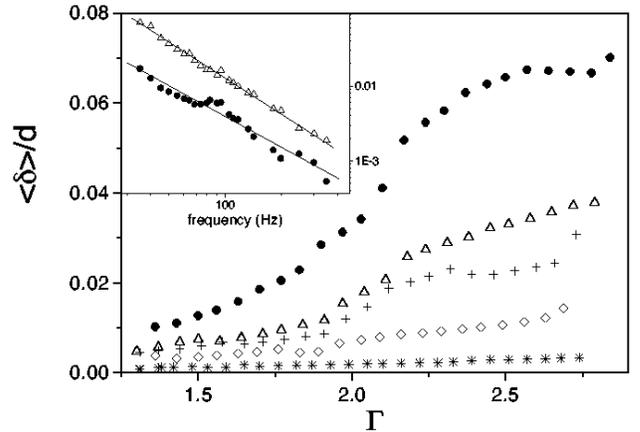}
\vspace{0.3cm}
\caption{\protect\small Average dilation, $\left < \delta \right >$, 
versus $\Gamma$  for 
various values of $f$; ($\bullet$) $f=40$ Hz, ($\triangle$) $f=55$ 
Hz,
($+$) $f=70$ Hz, ($\diamond$) $f=120$ Hz and ($\ast$) $f=250$
Hz. In the inset, log-log plot of  $\left < \delta \right >$ 
versus $f$ for $\Gamma=$ 1.5 ($\bullet$) and 2.4 ($\triangle$). 
The continuous lines show the fits $\left < \delta \right > \sim 
f^b$ 
for each $\Gamma$, with $b=-1.38 \pm 0.06$ and $b=-1.58 \pm 
0.03$ respectively.}
\label{fig8}
\end{figure}

Fig. \ref{fig8} presents  $\left < \delta \right >$ versus 
$\Gamma$ for several excitation frequencies. We observe  
in the low frequency regime that $\left < \delta \right >$ 
 presents a transition at $\Gamma \approx 2$.  In 
contrast, at high $f$ this transition has the tendency to 
disappear 
and we observe that $\left < \delta \right >$ is roughly a linear 
function of $\Gamma$.
Complementing the previous data, in the inset of Fig.  
\ref{fig8} we present $\left < \delta \right >$ versus $f$ for 
two values of $\Gamma$, below and above $\Gamma = 2$. 
The average dilation $\left < \delta \right >$ is a decreasing 
function of $f$.  We also present with continuous lines the fits 
$\left 
< \delta \right > \sim f^{b}$; we find $b = -1.38 \pm 0.06$ and $b 
= -1.58 \pm 0.03$ for $\Gamma=1.5$ and $\Gamma=2.4$ 
respectively. Similar behaviors are obtained for $\delta_b$, as a 
function of both $\Gamma$ and $f$ (See section IV).

To conclude, in this section we have introduced the peak shape of 
pressure from which we have obtained density profiles.  We 
have discussed these results in a qualitative way and found that 
dilation is a function of the vertical coordinate.  These results, 
which  
show that dilation is 
approximately constant in the bulk but increases sharply close to 
the free surface of the layer, are similar to the ones reported in 
previous works in one and two dimensions 
\cite{lcbrd94a,fermidirac}. We have also shown that at a low 
frequency of excitation, a critical value of $\Gamma \approx 2$ 
exists where an abrupt change in the layer dilation takes place.  
This transition will be discussed in detail in the next section.

\section{pressure and reflectivity measurements}

\begin{figure}[t!]
\centering
\leavevmode
\epsfxsize=8cm 
\epsfysize=4.9cm 
\epsfbox{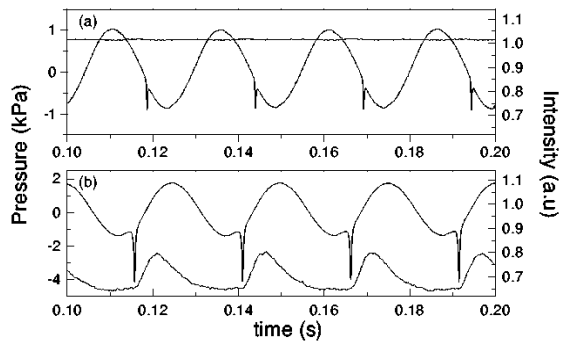}
\vspace{0.3cm}
\caption{\protect\small Time series obtained from pressure and 
intensity 
sensors for $f=40$ Hz; a) $\Gamma=1.2$, b) $\Gamma=2.3$.}
\label{fig9}
\end{figure}

Both the time-evolution of the pressure and the intensity are 
represented in Fig.~\ref{fig9} as a function of $\Gamma$ when $f$ is
in the low frequency regime.  For $\Gamma > 1$, the layer-plate
collision is 
always visible in the pressure signal as large peaks.  However, no 
trace of this collision is observed in the reflected light up to 
$\Gamma\sim {2}$.  Thus, for $1 < \Gamma < 2$ the layer is 
compact, indicating that the energy injected during the 
layer-plate collision is completely dissipated by the multiple 
collisions between the grains or by friction.  In contrast, for
$\Gamma>2$ the time mean value of the reflected light (DC 
component) exhibits a strong decrease that shows that the layer 
undergoes a transition from a compact to a dilated state.  
 A modulation in time (AC component) which oscillates at
the forcing frequency is also observed in the reflected light 
for
$\Gamma>2$.  This modulation is in phase with the pressure peak.  
Indeed, immediately 
after the pressure peak occurs, reflected 
light increases, indicating that the layer was dilated during the 
free flight, and that a small compression occurs due to the 
collision. Furthermore, when the layer 
takes off, surface dilation starts to increase (a decrease in 
reflected light).

We first emphasize that this increase in surface dilation 
is the result of 
the amplification of small differences in initial conditions, for 
the 
free flight of the grains located at the layer surface 
\cite{freexp}.  Such differences 
arise as a consequence of the random character of kinetic-energy 
injection; due to the random packing of grains, a layer plate 
collision  naturally induces velocity fluctuations in the layer.  As 
shown in Fig.~\ref{fig9} b for $\Gamma>2$,  the layer never 
reaches a compact state, which would correspond to a higher value 
of the 
reflected light (See Fig.~\ref{fig9} a). This result indicates 
that kinetic energy, 
injected into the internal degrees of freedom of the layer, has not 
been 
completely dissipated within the cycle.  To estimate the amount of 
energy not dissipated, we consider small changes of $I$ in time 
for early stages of the layer expansion.  We can then write   
$I(t)/I_{ref} \approx {1 - 2\Delta\delta_{s}(t)/(d + 
\delta_{s}(0))}$.  Here $\Delta\delta_{s}(t) = \delta_{s}(t) - 
\delta_{s}(0)$ and the reference
intensity $I_{ref}$ is taken when the layer starts to expand and 
corresponds to a finite dilation
 $\delta_{s}(0)$ at $t=0$.  From the slope of intensity versus 
time 
for early stages of the layer expansion we obtain 
a characteristic time $\tau$.  Dimensionally ${\tau}^{-1} \sim{ 
\Delta V_{to}/(d + \delta_{s}(0))}$,
where $\Delta V_{to}$ can be associated with the velocity 
fluctuations at the taking off time of the particles located
at the free surface of the layer.  Experimentally, $\tau$ is close 
to 
$1/27$ s and is almost independent 
of frequency.  Thus, our estimate shows that $\Delta 
V_{to}/V_{c}$ increases linearly with frequency like $2 \times  
10^{-4}f$, varying from 0.01 for low $f$ ($\sim{35}$ Hz) to 0.04 
for intermediate $f$ ($\sim{200}$ Hz). Here $V_{c}$ is the
layer-plate relative velocity at the collision calculated from the 
completely inelastic ball model.  
Therefore, our results indicate that for $2<\Gamma<2.8$, the ratio of residual 
energy to energy 
injection, 
$(\Delta V_{to}/V_{c})^2$, increases with $f$, implying that energy 
dissipation decreases with $f$.  At this stage, the important
feature of typical velocity fluctuations, or ``temperature", at the 
free surface of the layer arises:  we notice that although most 
of the energy is dissipated, a small amount of residual ``thermal energy" is 
enough to sustain surface dilation.

\begin{figure}[t!]
\centering
\leavevmode
\epsfxsize=8.5cm 
\epsfysize=6.2cm 
\epsfbox{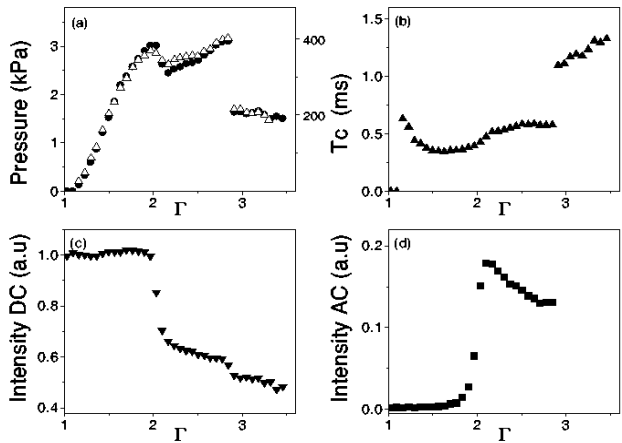}
\vspace{0.3cm}
\caption{\protect\small  Maximum pressure $P$, collision time 
$T_{c}$, DC and AC components of intensity versus 
$\Gamma$ for $f=40$ Hz. In (a) we present the fitting of 
the maximum pressure $P$  ($\bullet$) with $V_{c} / T_{c}$ 
($\triangle$), whose magnitude is presented in the right axis in units
of m/s$^2$.}
\label{fig10}
\end{figure}

We focus now on the transition from a compact to a dilated state 
suffered by the layer at $\Gamma\sim{2}$.  
Fig.~\ref{fig10} illustrates the pressure, $P$, the collision 
time, 
$T_c$, and the reflected intensity versus $\Gamma$ for $f = 40$ 
Hz. $P$ corresponds to the maximum pressure exerted on the 
plate during 
the collision and $T_c$ here is defined as the width of the 
pressure peak at a quarter of its height. We also present in Fig. 
\ref{fig10}a  the numerical fit of $P$ with 
$V_c/T_c$; it is evident that $P \sim V_c/T_c$, which is a natural 
consequence of the impulsive nature of the periodic forcing. At  
$\Gamma\sim {2}$, the DC component of the intensity exhibits a 
strong decrease while its AC component 
 abruptly increases. At the same value of $\Gamma$ there is a 
small decrease in the pressure peak, and a small increase in 
$T_c$, 
due to an increase in the bulk dilation in the layer. This 
decrease 
in pressure was already observed by P. Umbanhowar 
\cite{ums97}, and it is stronger for particles with higher 
restitution coefficients; however, no correlation with 
reflectivity 
measurements were made to investigate the state of the layer.  
We emphasize that the former transition occurs for $\Gamma < 
\Gamma_w$.  At the onset of 
surface waves ($\Gamma = \Gamma_w$), the pressure presents a strong 
decrease associated with the fact that layer-plate collision is 
spread 
out in time \cite{ums97} (Fig.~\ref{fig10} a, b).  

\begin{figure}[t!]
\centering
\leavevmode
\epsfxsize=8.5cm 
\epsfysize=6.1cm 
\epsfbox{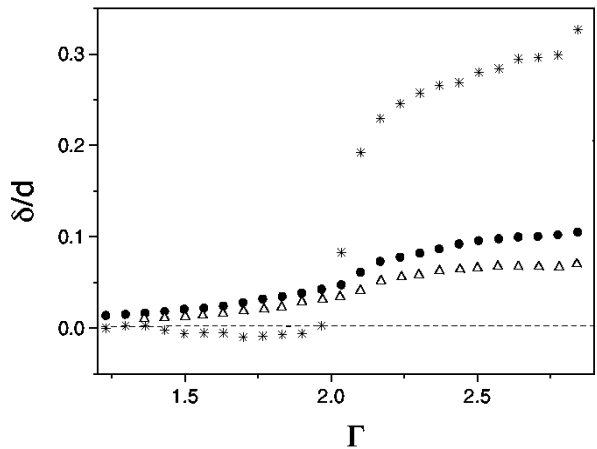}
\vspace{0.2cm}
\caption{\protect\small  Bulk dilation, $\delta_b$ 
($\bullet$), mean dilation ,$\left <\delta \right >$ ($\triangle$), 
and surface dilation, $\delta_s$ ($\ast$), versus 
$\Gamma$ for $f=40$ Hz.}
\label{fig11}
\end{figure}

Using the considerations introduced in the previous section, we 
calculate $\delta_b$ and $\delta_s$ as functions of  $\Gamma$.  
This is presented in Fig. \ref{fig11} for $f=40$ Hz.  To 
complete the 
data, we also present the average dilation $\left < \delta \right 
>$,
and we observe that 
the agreement with $\delta_b$ is fairly good.  
Similar to what we found for $\left < \delta \right >$ (fig. 
\ref{fig8}), we observe
that both $\delta_b$ and $\delta_s$ suffer transitions at $\Gamma 
\approx 2$.  
Curiously, for $\Gamma<2$ the bulk dilation is higher than the 
dilation in surface and $\delta_s$ takes negative values, which 
simply means that the layer surface reaches a state more compact 
than the initial one. This is related to the fact that the initial 
state, 
consistent with our experimental method \cite{init_cond}, is not the more 
compact accessible state of the layer.  
Therefore we say that for $\Gamma<2$ the state of the layer is 
solid-like, where the injected energy during the collision is 
completely dissipated.  However, in this regime, the available 
energy is enough to produce rearrangement of surface grains.  In 
all the cases tested, the maximum compaction was never larger 
than $2 \%$ of the initial density.  
Consequently, we associate the increase in $\delta_s$ and 
$\delta_b$, observed at low $f$ and $\Gamma>2$ to a solid-liquid 
type transition.  In the liquid phase, average dilation is large 
enough to allow particles to move with respect to each other.  For 
low $f$, at the critical value $\Gamma \approx{2}$, the injected 
energy rate 
becomes larger than the dissipation rate, and  the energy excess 
sustains the dilation in the granular layer.

\begin{figure}[t!]
\centering
\leavevmode
\epsfxsize=8.5cm 
\epsfysize=6.1cm 
\epsfbox{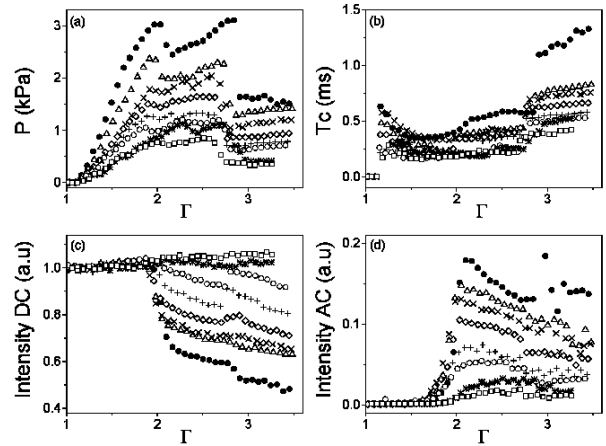}
\vspace{0.2cm}
\caption{\protect\small  Maximum pressure, $P$, collision time, 
$T_{c}$, and DC and AC 
components of intensity versus
$\Gamma$ for various $f$; $f$ = 40 ($\bullet$), 69 ($\triangle$), 
83 ($\times$), 111 ($\diamond$), 154 ($+$), 200 ($\circ$), 250 
($\ast$) and 350 ($\Box$) Hz.}
\label{fig12}
\end{figure}

\begin{figure}[t!]
\centering
\leavevmode
\epsfxsize=8.5cm 
\epsfysize=6.3cm 
\epsfbox{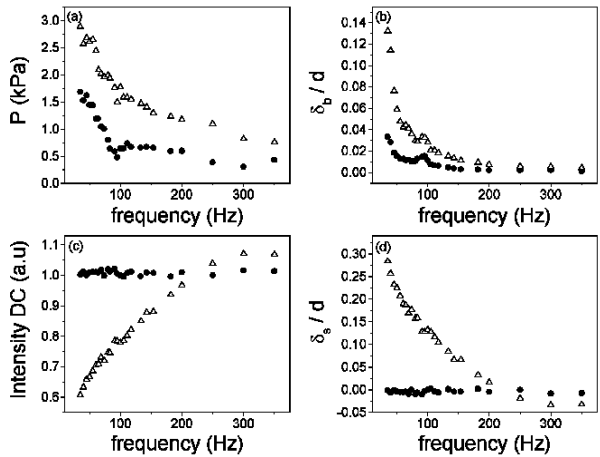}
\vspace{0.2cm}
\caption{\protect\small Maximum pressure $P$, bulk dilation 
$\delta_b$, DC component of intensity  and surface dilation 
$\delta_s$ 
versus $f$ for two constant values of $\Gamma$ below the onset 
of surface waves; $\Gamma$ = 1.5 ($\bullet$), 2.3 ($\triangle$). 
Bulk
dilation scales as $\delta_b \sim f^b$ with $b = -1.42 \pm 0.07$
($\bullet$)  and
$b = - 1.54 \pm 0.03$ ($\triangle$) respectively.}
\vspace{0.2cm}
\label{fig13}
\end{figure}

In Fig.~\ref{fig12} we include frequency dependence of the 
same quantities presented in Fig.~\ref{fig10}.  As opposed to 
the case of low $f$, at high $f$ and for $\Gamma>2$ the DC
component of $I$ increases slightly. This indicates that the layer 
surface has reached a state more compact than the initial one. It 
is very important to notice that the decrease in pressure associated 
with the wave instability is clearly observed over the entire range 
of frequencies (Fig.~\ref{fig12} a).
Complementing Fig.~\ref{fig12}, we present in Fig.~\ref{fig13} a and 
c the maximum
pressure and the DC component of intensity versus $f$ for two 
values of 
$\Gamma$, both smaller than the critical one for waves, with one right below 
and the other right above the fluidization transition.  Both the maximum 
pressure and  
the jump in reflectivity decrease 
as $f$ increases.

We also present the bulk dilation $\delta_{b}$ (See Fig.~\ref{fig13} 
b) calculated directly 
from the pressure through equation (\ref{p_vc2_delta}).
We find that $\delta_{b}$ scales as $f^b$ with $b = 
-1.42 \pm 0.07$ and $b = -1.54
\pm 0.03$ for $\Gamma=1.5$ and $ 2.3$ respectively. These values 
are in
very good agreement with those obtained for $\left < \delta \right 
>$
(See Fig. \ref{fig8}), and they provide a consistency test for 
both kinds of measurements. 
These results indicate that 
the relevant quantity is not only the ratio of the injected energy 
per particle to the potential energy required to rise a particle 
by a 
fraction of its diameter, in which case $\delta_{b}$ would vary as 
$1/f^{2}$, but also the dissipated energy.  We notice that our 
results
contrast with those obtained by Luding {\it et al} in numerical
simulations of a column of particles in the completely fluidized
regime, where the average dilation scales as $\left < \delta 
\right > \sim (Af)^2$,
which for $\Gamma$ constant becomes $\left < \delta \right > \sim
(\Gamma/f)^{2}$ \cite{lcbrd94a}. However, similar simulations 
done
by the same group for a two-dimensional layer \cite{lhb94} 
indicate
that the layer expansion scales as $h_{cm}-h_{cmo} \sim (Af)^{3/2} 
\sim
(\Gamma/f)^{3/2}$. Even though these results were also obtained 
in a completely fluidized regime (typically $\Gamma > 10$) we 
observe 
quite good agreement for the {\it scaling on $f$} for the 
expansion of the layer. 

On the other hand, information about the surface dilation,
$\delta_{s}$, versus $f$ is 
obtained using the intensity data as
$I/I_{0} \approx {d^{2}/(d + \delta_{s})^{2}}$ (See Fig.~\ref{fig13} 
d). 
For $f > 225$ Hz, $\delta_{s}$ takes negative values, which  
indicates that the surface layer is more compact than the initial 
one \cite{init_cond}.  However, in this regime, as shown in the
previous section, the 
bulk dilation increases with $\Gamma$ but remains very small,
$\delta / d \sim {10^{-3}}$.

\begin{figure}[t!]
\centering
\leavevmode
\epsfxsize=8.5cm 
\epsfysize=6.7cm 
\epsfbox{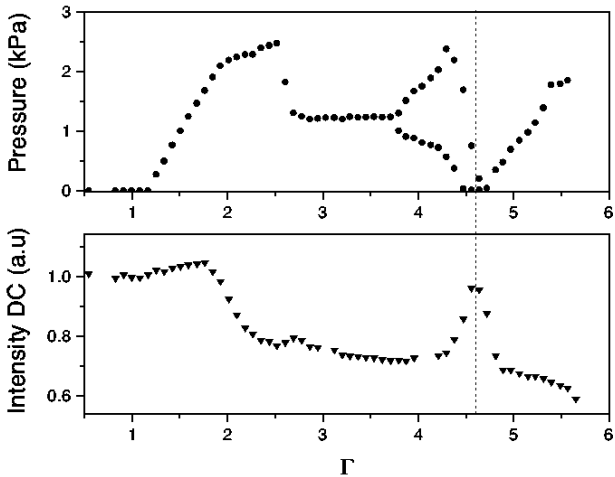}
\vspace{0.3cm}
\caption{\protect\small 
Maximum pressure  $P$ (a) and DC reflectivity-component (b) as a 
function of $\Gamma$ for $f=60$ Hz. The vertical line indicates 
the flat layer with kinks transition at which $V_c=0$.  At $\Gamma 
= 3.6$, as predicted by the ineslastic colliding ball model for 
the center of mass of the layer, a period doubling instability is 
observed.}
\label{fig14}
\end{figure}

Finally, let us mention another interesting 
transition which links to the flat with kinks instability reported 
in
previous works \cite{mus94}.  If we increase $\Gamma$ further we 
find
that a period doubling is achieved at $\Gamma \approx 3.6$, as
reported before \cite{dfl89,mus94}.  Next, for $\Gamma \approx 
4.6$, an inverse transition of the
liquid-solid type is detected. Fig. \ref{fig14} shows a typical
measurement of both reflected
intensity and maximum pressure for a wider range of $\Gamma$. We
detect that at $\Gamma \approx 4.6$, $P$ strongly 
decreases and the mean value of intensity strongly increases.  
Using the results discussed previously, both changes reflect a 
strong decrease of 
grain mobility. We notice that the critical value $\Gamma \approx
4.6$ corresponds to $V_c = 0$ in the completely inelastic ball 
model,
so in fact no energy is injected to the internal degrees of
freedom. We also conclude that no surface waves can be sustained 
in
this regime, except at the kink itself.  Indeed, the shear 
induced at
the kink by the flat parts oscillating out of phase is large and enough 
to
induce dilation.  At low frequencies, this
dilation is enough to allow hydrodynamics waves, as
those shown in figure \ref{fig1}c.  
As $\Gamma$ is 
increased further, the pressure increases and the reflectivity 
decreases. 
Thus, energy injection again becomes enough to sustain surface 
waves in 
the layer.  This fact is consistent with the existence of $f/4$ 
waves
reported previously \cite{mus94}. Notice that for $\Gamma = 4.6$ 
the
maximum velocity of the plate is $Aw \approx 12 $ cm/s; this 
confirms
that the relevant scale of velocity fluctuations is given by $V_c$ 
and not
$Aw$. 

\begin{figure}[t!]
\centering
\leavevmode
\epsfxsize=8cm 
\epsfysize=5.2cm 
\epsfbox{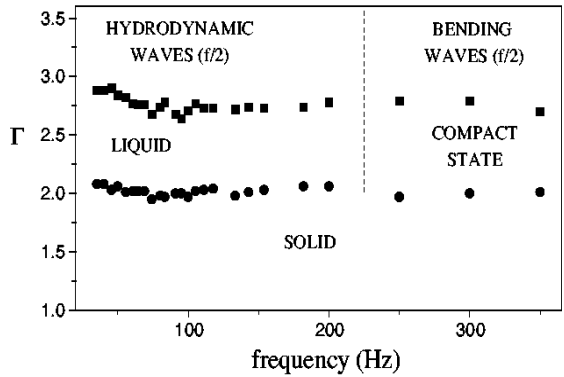}
\vspace{0.3cm}
\caption{\protect\small Phase diagram showing layer state and 
surface wave transitions. The vertical dashed line at $f=225$ Hz 
defines the frequency above which the layer strictly undergoes a 
transition to a state more compact at $\Gamma \sim 2$.}
\label{fig15}
\end{figure}

The experimental results presented above can be summarized as 
follows.  Depending on the excitation frequency 
we observe different kind of states and waves.  At low frequency, 
bulk and surface dilation present  strong increases which are 
associated with a fluidization transition.  Surface waves observed 
in 
this regime involve large relative motion between particles and 
are therefore considered as the hydrodynamic modes of the layer.  
At intermediate 
$f$, although the injected energy is small, we still observe a 
 decrease in reflectivity as a function of $\Gamma$. In 
this regime, as shown in Fig.~\ref{fig13} b, $\delta_b$ 
 is too small ($\delta_b/d<0.1$) to allow 
motion between the particles \cite{reynolds}. Therefore, at the 
critical 
$\Gamma$, the 
decrease of reflectivity is the signature of particles fluctuating 
around their positions at the free surface of the layer. We 
associate this decrease with a heating up of the solid phase. For 
higher 
frequencies, the layer undergoes a compaction transition which is
detected by the increase in surface density.  Below and above this 
transition the local bulk dilation $\delta_b / d$ is even smaller 
and is of order $0.005$, implying that the mobility in both the bulk and the 
free surface is 
completely suppressed.
Thus, very 
low amplitude surface waves detected in the compaction regime, 
 by the strong decrease in the maximum pressure,
 must correspond to excitations in which the layer is slightly 
modulated in time and space.  We will see in the 
following section that these waves are bending waves, 
associated with the ability of the compact layer to deform. 
Finally, fig.~\ref{fig15} presents the phase diagram for the granular 
layer: phase 
boundaries separating the various states and surface waves have 
been obtained from the data in Fig.~\ref{fig12}.  The layer 
state and 
surface wave transitions occur for approximately constant values 
of $\Gamma$ independent of $f$.

\section{Low amplitude waves: bending waves}

To check the existence of these waves we have performed a set of 
experiments in a two-dimensional granular layer.  The advantage 
of using a two-dimensional system is that this allows us to obtain  
side views of the waves.  This is of particular importance since 
bending 
waves are difficult to visualize as they have amplitudes of the 
order of fifty percent of a particle diameter.  

In order to observe such small amplitudes we have used 
photoelastic cylinders of $6$ mm in diameter and $6.35$ mm in 
length. Typical excitation frequencies are about $40$ Hz, which 
for 
this system 
are in the high frequency regime.  We recall that this regime 
occurs for 
frequencies much larger than the 
frequency crossover, $f\sim{\sqrt{g/d}}$.  When varying the 
deepness of the layer (number of particles = $N$), this scaling 
becomes 
$f \sim{\sqrt{g/Nd}}$ \cite{bizon}.
In the case of bronze 
particles of $d \approx 0.12$ mm, bending waves are detected for 
$f > 225$ 
Hz.  This tells us that these waves should be observed at 
$f \approx{40}$ Hz, for $d=6$ mm  and a layer about ten particle 
diameters thick.  With these parameters, for 
$\Gamma \sim 3.5$, the amplitude of waves should be of the order of 
$1$ mm. 

\begin{figure}[t!]
\centering
\leavevmode
\epsfxsize=6.3cm 
\epsfysize=14cm 
\epsfbox{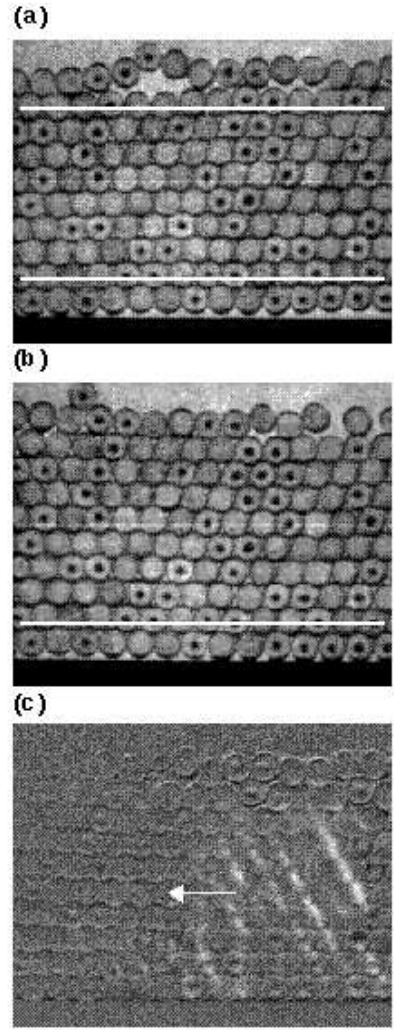}
\vspace{0.3cm}
\caption{\protect\small Typical snapshots of bending waves for $f 
=40$ Hz and $\Gamma = 3.6$. In (a) and (b) the white horizontal 
lines show that the layer bends itself about half a diameter. Both 
states alternate in time with the frequency of oscillation. In (c) 
we 
present the difference of two images during a collision. The 
bright 
zones correspond to regions under high stress. In this case the
compression front travels to the left.}
\label{fig16}
\end{figure}

Typical snapshots of two stages of bending waves at 
$\Gamma=3.5$, $f=40$ Hz and $N=10$ are presented in figures 
\ref{fig16}a and 
\ref{fig16}b.  We observe that the layer slightly bends with 
respect to the 
horizontal.  Similar to what occurs for low frequency waves, 
this 
modulation alternates in time at half of the frequency forcing 
(See below).
However, in this case,  
the wavelength of the modulation is about a layer 
thickness and is also nearly independent of $f$.  Some particles 
are marked with black spots which allows us to 
distinguish them and follow their trajectories (See Fig. 
\ref{fig16}a, b).  As expected, the mobility of the particles is 
very low.  Only at the surface layer will some particles move with respect 
to each other over distances of the order of the driving amplitude.  In 
the bulk this 
motion is completely suppressed.

Additional support for these low amplitude waves is provided in 
Fig. \ref{fig16}c in which a compaction front that moves 
laterally is observed as a bright zone. This image results from 
the
difference of two consecutive snapshots (period of acquisition is
about 0.8 ms).  At a compression zone, the 
vertical stress in the layer is high so the light is transmitted. 
Thus, the
difference of images mainly shows the zones under stress. In the
case presented the compression front moves from right to left and 
the
collision occurs very near the right boundary of the image. In the
next cycle the collision occurs near the left boundary and the
compression front travels to the right.  With a
wider view of the layer we observe that in fact two compression 
fronts
are created from
each collision point; one front travels in each direction. This 
kind
of visualization allows us to safely say that these parametric 
waves
are also subharmonic, i.e. their frequency is $f/2$. Finally, from 
the
images we estimate the velocity
of the compression front of the order of a few meters per
second. This value is very high compared to the estimated velocity 
at the
collision $V_c \approx 0.25 $ m/s; this indicates the existence of
contact arcs (see the contact lines in Fig. \ref{fig16}c) so the
layer is almost not dilated.

\section{Conclusions}

In conclusion, depending on the excitation frequency, 
we observe different kinds of states and waves.  Thus, our 
experimental results reveal the existence of a 
solid-liquid transition that precedes subharmonic wave 
instability.  
Hydrodynamic surface waves can be then considered as the 
natural 
excitations existing in a fluidized granular layer. In contrast, 
very 
low amplitude surface waves detected in the compaction regime 
correspond to excitations in which the layer slightly bends
 alternatively in time and space.  We have seen in the 
previous section that these waves are associated with the 
compact  character of the layer.  The layer states 
observed 
here and
surface wave transitions occur for approximately constant values 
of $\Gamma$ independent on $f$.

Additional experimental evidence of the fluidization transition 
experienced by the layer at $\Gamma\sim{2}$ can be found in 
reference \cite{brennen}  (See for instance, Fig. 6  in ref. 
\cite{brennen}).  Although they used 
rather large particles at small forcing frequencies, we find that their
experiments are actually in the low frequency regime after estimating 
the frequency crossover.  Independent experimental evidence 
for the compaction transition is also found in previous works.  
For instance, a strong increase in granular density was found, 
close 
to $\Gamma\sim{1.9}$, in several experiments on large columns of 
grains submitted to ``taps'' of a single cycle of vibration
\cite{knight}.  In 
this case, applying the scaling to the frequency crossover, we 
found 
consistently that these experiments correspond to the high 
frequency limit.  Therefore, 
we conclude that both fluidization as well as compaction 
transitions are well established experimentally.  However, it is 
still 
unclear what mechanisms dominate these transitions 
and why they arise at a constant value of $\Gamma$.  We can only 
safely say that at low energy injection rates, or equivalently at 
small plate amplitudes with respect to the thickness of the layer, 
the transition will be of the compaction type.  In the opposite 
case of high energy injections rates, this transition will be of  
the fluidization type.

Also, we have clearly shown that the relevant scale of velocity 
fluctuations is
given by $V_c$ and not $Aw$. Then, via the intensity measurements 
we deduced
that the dissipation decreases as $f$ is increased; this is 
equivalent
to state that the dissipative effects increase with velocity 
fluctuations.
A possible cause of this is the
dependence on the velocity of the restitution coefficient: it
has been recently shown that $1 - \epsilon \sim v^\alpha$, where $v$
is the relative normal velocity at the colision and $\alpha$ a positive
number \cite{flfc98}. Nevertheless, as the dilation is reduced we expect 
that
the friction between grains will become an important dissipative
mechanism. 

Finally, the importance of the bulk and surface dilation 
measurements is
that they provide a complementary way, with respect to the granular
temperature, to explore the excitation of the internal degrees of freedom
of a vibrated layer. For the dilation in the bulk
we have found that, for a constant value of $\Gamma$, it is a 
decreasing function of $f$ of the form
$1/f^b$, with $b \approx 3/2$. This agrees with previous 
simulations
of a two dimensional layer \cite{lhb94}. It is clear that the
deviation from the expected value $b = 2$ is due to dissipative
effects, but the exact numerical value seems to depend on
$\Gamma$. 

It is a pleasure to acknowledge to Paul Umbanhowar and Enrique 
Tirapegui for many enlightening discussions and to
Satish Kumar for useful comments on the manuscript.
This work was supported by Fondecyt Grant $N_{0} 1970682$, 
Catedra Presidencial en Ciencias and Dicyt USACH.

\end{multicols}


\begin{thebibliography}{10}

\bibitem{jnb96} For recent reviews, see H.M Jaeger, S.R. Nagel, 
and 
R.P. Behringer, Phys. Today {\bf 49}, No 4, 32 (1996); Rev. Mod. 
Phys. {\bf 68}, 1259 (1996).


\bibitem{mus94} F. Melo, P. Umbanhowar and H. L. Swinney, Phys. 
Rev. Lett. {\bf 72}, 172 (1994).  F. Melo, P. Umbanhowar and H. L. 
Swinney, Phys. Rev. Lett. {\bf 75}, 3838 (1995).

\bibitem{haff} It is well known that the mobility in a granular 
material 
depends strongly on the energy injection rate.  See for instance 
P. 
K. Haff, J. Fluid Mech. {\bf 134}, 401 (1983).

\bibitem{ums96} P. Umbanhowar, F. Melo and H. L. Swinney. 
Nature {\bf 382}, 793 (1996).


\bibitem{mm98} N. Mujica and F. Melo, Phys. Rev. Lett. {\bf 80} 
5121 (1998).

\bibitem{pdb95} H.K. Pak, E. Van Doorn, and R.P. Behringer, Phys. 
Rev. Lett. {\bf 74}, 4643 (1995).

\bibitem{init_cond}  $I_{0}$ is obtained after a "fluidization" of 
the layer, for many cycles, at $\Gamma=2.4$ and $f=40$ Hz. Before 
each measurement we set the initial state of the layer by this 
procedure.

\bibitem{ums97} P. Umbanhowar, Ph.D. Thesis, University of 
Texas at Austin, (1996).  P. Umbanhowar, F. Melo and H. L. 
Swinney, (unpublished).


\bibitem{falcon} E. Falcon, C. Laroche, S. Fauve and C. Coste, 
Eur. 
Phys. J. B {\bf 5}, 111 (1998).

\bibitem{landau_elasticity} L.D. Landau and E.M. Lifshitz, Section 
9, 
{\it Theory of Elasticity}, $3^{rd}$ Edition, Pergamon Press 
(1986).

\bibitem{tiempos} Both times $T_q$ and $\tau_1$ depend on the 
velocity of collision as $V_c^{-1/5}$ and the estimates in the 
text 
correspond to the lowest values of $V_c$ achieved in our 
experiments.

\bibitem{gsg96} A. Goldshtein, M. Shapiro and C. Gutfinger,
J. Fluid. Mech. {\bf 316}, 29 (1996).

\bibitem{lcbrd94a} S. Luding, E. Clement, A. Blumen, J. 
Rajchenbach, 
and J. Duran, Phys. Rev. {\bf E, 49}, 1634 (1994).

\bibitem{fermidirac}  E. Clement and J. Rajchenbach, Europhys. 
Lett., 
{\bf 16}, 139 (1991), J.A.C Gallas, H. Herrmann, and S. 
Sokolowski, 
Physica (Amsterdam) {\bf 189A}, 437 (1992).


\bibitem{freexp}  Simple numerical simulations show that the 
experimental variation of intensity, during the free flight of the 
layer, is reproduced by assuming small velocity fluctuations at 
the 
take off time for the particles located at the free surface.  
Such 
velocity fluctuations are very close to the ones estimated in the 
text.
N. Mujica and F. Melo, (unpublished). 

\bibitem{lhb94} S. Luding, H. Herrmann and A. Blumen, Phys. Rev
E. {\bf 50}, 3100 (1994).


\bibitem{dfl89} S. Douady, S. Fauve and C. Laroche, Europhys. 
Lett.
{\bf 8}, 621 (1989).

\bibitem{reynolds} O. Reynolds, Philos. Mag. Ser. 5, {\bf 20}, 469 
(1885).


\bibitem{bizon} C. Bizon, M.D. Shattuck, J.B. Swift, W.D. 
McCormick 
and H. Swinney, Phys. Rev. Lett. {\bf 80} 57 (1998).

\bibitem{brennen} C.E. Brennen, S. Ghosh and C.R. Wassgren, J. 
Appl. Mech. {\bf 63}, 156 (1996).

\bibitem{knight} J. B. Knight, C. G. Fandrich, Chun Ning Lau, H. 
M. 
Jaeger, and S. R. Nagel, Phys. Rev. {\bf E 51}, 3957 (1995).

\bibitem{flfc98} E. Falcon, C. Laroche, S. Fauve and C. Coste,
Eur. Phys. J. B. {\bf 3}, 45 (1998).

 
\end{thebibliography}
\end{document}